\documentclass[a4paper,
               keeplastbox,   
               ]{jacow}
\usepackage{pdfpages,multirow,ragged2e} %
\usepackage{todonotes}
\makeatletter%
	\ifboolexpr{bool{xetex}}
	 {\renewcommand{\Gin@extensions}{.pdf,%
	                    .png,.jpg,.bmp,.pict,.tif,.psd,.mac,.sga,.tga,.gif,%
	                    .eps,.ps,%
	                    }}{}%
\renewcommand\@seccntformat[1]{}
\makeatother
\ifboolexpr{bool{xetex} or bool{luatex}} 
{}                                      
 {\usepackage[utf8]{inputenc}}           

\usepackage[USenglish]{babel}

\providecommand{\keywords}[1]{\textbf{\textit{Index terms---}} #1}

\begin{document}

\title{Privacy in Open Search: A Review of Challenges and Solutions}

\author{Samuel Sousa\thanks{ssousa@know-center.at}\textsuperscript{1}, Roman Kern\textsuperscript{1}, Know-Center GmbH, Graz, Austria\\
\textsuperscript{1}also at ISDS, Graz University of Technology, Graz, Austria\\
Christian Guetl, ISDS, Graz University of Technology, Graz, Austria\\ }

	
\maketitle

\begin{abstract}

Privacy is of worldwide concern regarding activities and processes that include sensitive data.
For this reason, many countries and territories have been recently approving regulations controlling the extent to which organizations may exploit data provided by people.
Artificial intelligence areas, such as machine learning and natural language processing, have already successfully employed privacy-preserving mechanisms in order to safeguard data privacy in a vast number of applications.
Information retrieval (IR) is likewise prone to privacy threats, such as attacks and unintended disclosures of documents and search history, which may cripple the security of users and be penalized by data protection laws.
This work aims at highlighting and discussing open challenges for privacy in the recent literature of IR, focusing on tasks featuring user-generated text data.
Our contribution is threefold:
firstly, we present an overview of privacy threats to IR tasks;
secondly, we discuss applicable privacy-preserving mechanisms which may be employed in solutions to restrain privacy hazards;
finally, we bring insights on the tradeoffs between privacy preservation and utility performance for IR tasks.


\end{abstract}


\section{Introduction}\label{sec:introduction}

Data is the cornerstone of many research fields, as well as the source of information used by professionals, such as journalists, statisticians, and policymakers~\cite{brickley2019google}. 
At the pace of the Internet popularization, the number of data publishers rose to the thousands~\cite{magazine2017landscape}, encompassing university library files, government open data, Wikipedia articles, social media platforms, commercial data providers, digital data markets, scientific data repositories, among many others.
Accessing these data sources is crucial for solving the reproducibility issues of scientific results and improving the means for data journalists to obtain reliable information~\cite{brickley2019google}.
Furthermore, the idea of open science relies on sharing research data and materials for third-parties to reuse and reproduce experiments, assuring the trustworthiness of the research process~\cite{magazine2017landscape}. 
Some companies also need to have their products openly visible on Internet for their business to keep going, e.g., e-commerce platforms.
However, privacy issues may keep some kinds of data from being released in order to avoid the exposure of confidential information.
As an example, user-generated data, such as user behavior, search interests, and search profiles, demands anonymization steps prior to its release.
Otherwise, it should remain unrevealed.
Other privacy issue regards the level playing field for search services since those which monitor the intents of users can eventually obtain benefits out of it. 

Privacy is a concept related to limiting the extent of information an individual is willing to share~\cite{westin1968privacy}.
Many countries consider privacy as a right protected by law.
For instance, the General Data Protection Regulation (GDPR)~\cite{EUdataregulations2018}, which entered into force in May 2018 in the European Union (EU), draws guidelines for the collection, transfer, storage, management, and deletion of personal data within the member states of the economic bloc.
GDPR grants the residents of the EU the control over their personal data.
Therefore, any processing activities over personal data must comply with the regulation and provide protection measures.
In case of data breaches or noncompliance with the legal principles, penalties and fines are applicable.
Since 2018, data protection bills have also been passed in several countries worldwide.

In the recent past years, the preservation of privacy has been gaining attention in the field of information retrieval (IR)~\cite{zhang2018tagvisor,zhu2017differentially,weng2014privacy,tamine2018evaluation}.
Several privacy-preserving mechanisms have been developed to safeguard personal data from threats, as attacks, disclosures, and unintended usages.
Some of these mechanisms, such as encryption~\cite{gentry2009fully,cash2015leakage}, differential privacy (DP)~\cite{dwork2008differential}, multi-party computation (MPC)~\cite{feng2020securenlp}, and federated learning (FL)~\cite{konevcny2016federated,chen2019federated}, are implemented alongside models for enabling applications to safeguard data privacy.
There are also some attacking methods that aim at retrieving data samples used to train models, mostly neural networks, which can be seen as a threat or a safety checker, depending on the attacker's intention.
Furthermore, some privacy-preserving tools may present computational overheads or performance reductions, which call attention to privacy-utility trade-offs.

Searches on personal data often present privacy risks from both data provider and model sides.
There is a need for open data, due to scientific, governmental, and press reasons~\cite{brickley2019google,magazine2017landscape}, however data generators and IR systems users cannot have their identities and personal data exposed.
This work aims, therefore, at pointing out open challenges with regards to privacy in IR tasks, as well as reviewing appropriate privacy-preserving methods to safeguard personal data or model.
We focus primarily on tasks featuring text data since the private content of data in written format may be presented explicitly, as a person's name or an ID number, or implicitly like a the gender of a person which can be inferred from the text, based on gendered words like profession terms.

Our contributions comprise a summary of research directions on privacy for IR tasks alongside adequate privacy-preserving methods for the privacy risks these tasks may present.
Problems that put privacy at risk are also discussed.
Moreover, we discuss how privacy-preserving methods can influence the results of IR tasks.
Finally, we provide the readers with essential understanding of privacy-related issues and privacy-preserving methods for IR.

This paper is organized as follows.
Firstly, we review recent works in the literature of privacy in section \textit{Related Work}.
Secondly, we describe open challenges and solutions for privacy preservation in open search tasks in section \textit{Privacy Challenges and Solutions}.
Further, we discuss our results in the section \textit{Discussion}.
Finally, our contributions and upcoming works are brought into context in the section \textit{Conclusion and Future Works}.

\section{Related Work}\label{sec:background-and-related-work}

Data privacy is a large concern within IR~\cite{zhu2017differentially,weng2014privacy,tamine2018evaluation} since privacy threats and risks often arise from searches on private data~\cite{zhu2017differentially} and tasks which involve the search behavior or intent of users~\cite{orso2017overlaying,weng2014privacy,tamine2018evaluation}.
Examples of such issues include private data publishing~\cite{zhu2017differentially}, string searches~\cite{riazi2017prisearch}, user's context~\cite{tamine2018evaluation}, information sharing of web search logs~\cite{mivule2017data}, interactive search~\cite{orso2017overlaying}, and the information selection behavior of users~\cite{orso2017overlaying}.
Additional privacy threats include revealing private data from IR systems that compute distributed information, performing face recognition without the consent of the people whose faces are captured, and so on.
Therefore, a large number of real-world IR systems can be prone to breaches of personal data, unintended disclosures of private information, and penalties established by data protection regulations.

Nowadays, many organizations collect and process personal data, such as governments, companies, and search service providers.
As a consequence, the volume of collected data has increased alongside the risks related to breaches of sensitive information from these datasets.
Zhu et al.~\cite{zhu2017differentially} survey DP applications for data publishing and data analysis.
The authors define data publishing as publicly sharing data itself or the result of queries, whereas data analysis consists on releasing data models to the public.
In both scenarios, DP can offer privacy guarantees resistant against attacks and mathematically provable. 
Riazi et al.~\cite{riazi2017prisearch} come up with a mathematically provable mechanism for privacy preservation in IR tasks.
The authors implement a methodology for two-party string search based on the Yao's Garbled Circuit protocol.
For instance, two users can hold queries and data for string search simultaneously, whereas they both aim at keeping these search components private without relying on a third party like a trusted server.
Therefore, the proposed protocol converts the search algorithm into a Boolean circuit that evaluates the private queries and texts.
Tamine and Daoud~\cite{tamine2018evaluation} survey methodologies and metrics for context IR evaluation, specifying the impact of data privacy towards the evaluation design.

User search behavior can also be seen as private information since queries and search result's selection can disclose demographic attributes, preferences, political views, etc.
Orso et al.\cite{orso2017overlaying} investigate the role of user search behavior and information selection in order to understand which layers of social information, namely personal preferences, tags combined to personal preferences, or tags and social ratings combined with personal preferences, can enhance search efficiency.
The authors found empirical evidence that personalized preferences and social ratings make it easier for users to select information without external sources.
In this study, the publicly available Yelp dataset\footnote{\url{https://www.yelp.com/dataset}.} was used.
Therefore, privacy aspects that would arise from this use case in a real-world scenario, as the compliance with data protection regulations, were not approached.
Sharing search log records is a process ruled by data protection laws, which demand anonymization techniques to be applied before the data leaves the servers it is stored on.
Mivule et al.~\cite{mivule2017data} introduce an heuristic for privacy preservation of individual web search log records based on swapping.
Firstly, an individual has a set of logs $\mathcal{A}$.
Secondly, the records in $\mathcal{A}$ are switched within this set.
Finally, the swapped files from $\mathcal{A}$ are switched again using records of a set $\mathcal{B}$.
This heuristic is efficient when it comes to preventing an attacker from tracing the issuer of a search query.
Additional privacy-preserving IR solutions include homomorphic encryption (HE)\cite{el2021papir}, DP~\cite{yang2017differential}, and hashing~\cite{weng2014privacy}.

Our work focuses on identifying privacy risks across IR tasks, alongside suggesting privacy-preserving mechanisms that have the potential to suit the needs for privacy protection.
We briefly introduce the task description followed by the privacy risks associated with both data and IR model.

\section{Privacy Challenges and Solutions}\label{sec:privacy-challenges-and-solutions}

This section firstly introduces open challenges with regards to privacy in IR tasks.
We survey recently published works for ten IR tasks, highlighting their privacy issues.
Afterwards, we overview privacy-preserving methods that are suitable options to address these problems.

\subsection{Privacy Challenges}\label{sub:challenges}

\subsubsection{Search tasks}

\paragraph{Ad-hoc search.}
Modern search engines often rely on bag-of-words models to represent documents and search queries. 
Consequently, accurate quantification of context-specific term importance in documents is a tricky problem since term's context is often not captured by these models. 
When it comes to data privacy, bag-of-words models pose risks related to data re-identification.
For instance, some terms in the vocabulary of the model may refer to personal identities or private attributes, such as age, gender, location, and demographic information, which allow a de-identified document to be re-identified.
Dai and Callan~\cite{dai2020context} propose a novel document term weighting framework which preserves word context, using BERT~\cite{devlin2019bert} embeddings to extract document representations.
Although keeping document's contextual information, BERT embeddings can also encode private information and suffer attacks as model inversion, reverse engineering, and membership inference.

\paragraph{Query expansion.}

Query expansion (QE) helps users of IR applications to find more relevant information by expanding the search queries, hence increasing recall.
For instance, synonyms and hypernyms of terms in the questions for question answering (QA) are used to rise the likelihood of matching sentences stating the most appropriate candidate answer~\cite{esposito2020hybrid,carpineto2012survey}.
However, in some QA scenarios the number of retrieved sentences may be small, and then mismatch the intents of the user who queries the QA system~\cite{carpineto2012survey}.
Common privacy threats in applications which use QE regard disclosing query content or private information in the documents in which the sentences are extracted, such as names and locations of people.

\paragraph{Feature extraction for ranking.}

In the context of ad-hoc search, document ranking consists on returning a ranked list of documents from a large collection based on the assumed information need expressed by a search query, maximizing some ranking metric like average precision~\cite{nogueira2020document}.
Ranking is a cornerstone for IR systems and search engines, which can also be performed by machine learning (ML) classification models~\cite{pandey2018linear,nogueira2020document}.
Therefore, in order to reduce computation time, boost learning results, and prevent overfitting for those models, feature extraction techniques can be implemented to better represent documents.
Pandey et al.~\cite{pandey2018linear} come up with a method for representing documents as matrices with reduced dimensions when compared to the original document representations.
Such representations are useful for improving the results of ranking algorithms.
However, feature extraction models can have its original training samples disclosed by attacks of model inversion, membership inference, or reverse engineering.

\paragraph{Online learning for ranking.}

A critical drawback of ranking models regards the hardness to obtain labelled data for model training.
Thus, Zhuang and Zuccon~\cite{zhuang2020counterfactual} propose a counterfactual learning to rank method based on logs of user interaction collected from the ranking model in production.
In a real-world setting, users would be able to confirm the effectiveness of the ranking model by clicking on the results.
In the experimental evaluation, the authors use publicly available datasets and generate user clicks automatically with a cascade click model~\cite{chuklin2015click}.
This scenario can pose privacy threats to personal data collection if the user clicks are collected from actual system users.
Therefore, regulations for personal data collection would be applicable alongside the need for privacy-preserving methods.

\paragraph{Query composition.}

ML algorithms can be used to predict query properties like answer size, run-time, and error class~\cite{zolaktaf2020facilitating}.
These algorithms can therefore be prone to unintended memorization of query content alongside the attacks which aim at disclosing training document samples.
Another privacy threat regards the location of the ML model, e.g., on a cloud server, since computation parties sometimes may not be trusted, or the communication channels for transferring data or model updates may allow eavesdropping attacks to take place.



\subsubsection{Healthcare tasks}

\paragraph{Healthcare data tasks.}

Electronic medical records or electronic health records are digital documents, in which medical staff inputs patient data, including personal information, health condition, disease diagnosis, medication, etc.~\cite{sun2018data}.
This type of document has the advantages of easy storage, transfer, sharing, and deletion.
However, healthcare data is inherently private due to the sensitivity of its content.
Therefore data protection regulations, as the EU's GDPR~\cite{EUdataregulations2018}, prevents publicly releasing such data for research activities and public searches, without the explicit consent of data owners and the application of data anonymization methods.
In the context of IR, medical applications have to safeguard medical data from queries by malicious users or computation parties, e.g., corrupted servers.

\subsubsection{Social media tasks}

\paragraph{Opinion mining.}

Many online data sources contain opinions, which can be classified with regards to their polarity.
For instance, Nguyen and Nguyen~\cite{nguyen2018multilingual} predict sentiment polarity on YouTube\footnote{\url{http://youtube.com}.} comments in English and Italian languages, using a DL model based on a Bidirectional Long Short-Term Memory architecture.
Tasks performed over user-generated data pose privacy risks of unintended data memorization by the neural network, as well as sensitivity to attacks that aim at retrieving the model training samples.
Therefore, there is a room for privacy-preserving mechanisms that prevent such privacy issues.

\paragraph{Advisor for hashtag sharing.}

On social media platforms, users often put their privacy at risk unconsciously by releasing details of their personal lives publicly, revealing their exact location, and posting political or societal views.
Furthermore, some features of such platforms like hashtags may induce privacy threats, mainly related to location, since attacking models can easily predict precise user location from the hashtags they post online~\cite{zhang2018tagvisor}.
This situation can be prevented by privacy-preserving mechanisms of data obfuscation or de-identification, which get rid of privacy-sensitive information before any sharing step by the user.
Moreover, social media data is frequently protected by data protection regulations, then, ensuring data privacy to be safeguarded.

\paragraph{Social media profile linking.}

User identity linking is the task of connecting accounts owned by the same user across different social media platforms~\cite{hadgu2020learn2link}.
For instance, a user can have profile simultaneously on TikTok\footnote{\url{https://www.tiktok.com/}.}, Twitter\footnote{\url{https://twitter.com/}.}, Instagram\footnote{\url{https://www.instagram.com/}.}, and so on.
However, more relevant advertisement can be suggested to this user by linking these different profiles.
This scenario, therefore, presents privacy threats related to monitoring user behavior online besides linking profiles where the user uses a pseudonym to remain anonymous.

\subsubsection{Recommendation tasks}

\paragraph{Recommender systems.}

Recommender systems aim at predicting the preferences of users based on their interest, making more effective use of information~\cite{wang2015collaborative}.
However, user interest and their search history should be preserved from leakages, hence these are pieces of private information.
A recent application of recommender systems is news recommendation.
Qi et al.~\cite{qi-etal-2020-privacy} propose a framework to recommend news to users in distributed computation scenarios, e.g., smartphones running a mobile application.
This frameworks computes updates locally on each distributed device separately and, then, sends these local updates to a centralized server which updates the global model by aggregating the local updates.
Finally, the updated global model is distributed over the distributed devices to news recommendation and successive updates.
This is an outstanding application of FL, which prevents threats of data leakage from the distributed devices.
However, this model still can suffer from unintended memorization of user's behavior, preferences, and search history.

\subsection{Solutions}\label{sub:solutions}

Many privacy-preserving methods have been proposed for the sake of preventing attacks and unintended data breaches.
In this subsection, we review computational techniques for privacy preservation, which can be  conveniently integrated into IR tasks.
We group these techniques according to their technical aspects into cryptographic approaches and ML-based applications.

\subsubsection{Cryptographic approaches}

Cryptographic protocols have been extensively used to protect private data when sharing activities are not advised.
In a nutshell, encryption consists on the application of a function over data in raw format, which is referred as `\textit{plaintext}'.
This function results on an output called `\textit{cyphertext}', which inhibits the identification of its original format or content.

\paragraph{Encryption.}
Homomorphic encryption (HE) is a form of encryption that allows arithmetic operations to be computed over cyphertexts without the need for decryption \cite{gentry2009fully}. 
For instance, ML models can be used for inference on encrypted data, whereas the results are still consistent.
Encryption schemes that implement HE are particularly useful for scenarios in which data transfers and centralized storage occur, e.g., cloud servers, alongside the lack of trust.
However, it is worth to mention the computational overheads that FHE often leads to, hence its use becomes prohibitive for devices with small memory capacity.

IR models can also be privacy-preserving by the use of searchable encryption (SE).
This technique encrypts document collections enabling the data owner to delegate search capabilities, whereas the server or a service provider, like a search engine, does not demand decryption~\cite{cash2015leakage}.
Therefore, the so-called `honest-but-curious' server can provide searches, whereas the content of the stored data and the input queries is preserved~\cite{liu2020mitigating}. 
On the other hand, semantic relations between words and documents may be lost in the encrypted forms, so that decaying search results~\cite{dai2019efficient}. 
SE usually encompasses algorithms for the steps of key generation, encryption, token generation, and search~\cite{liu2020mitigating}. 
Finally, the main threat this encryption scheme faces is related to keyword inference attacks that aim at recovering the content of encrypted keywords.

\paragraph{Multi-party computation.}

Document collections may be stored in distributed search system hosted by other companies or even distributed across members of a broad community.
Therefore, when the exchange of documents among the members of these computation scenario is not an advisable option, multi-party computation (MPC) can be successfully used.
MPC is a cryptographic primitive that computes aggregated functions over multiple sources of data, which cannot be revealed~\cite{feng2020securenlp}. 
Formally, MPC assumes a set of inputs $\{x_1,x_2,\dots,x_n\}$ so that each party $P_i$ will store $x_i$ and agree to compute $y = \hat{f}(x_1,x_2,\dots,x_n)$, in which $y$ is the output information to be released, and $\hat{f}$ is the agreed function on the entire input set~\cite{cramer_damgard_nielsen_2015}. 
The input set may be composed of keywords, documents, medical records, etc.

\subsubsection{Differential privacy}
DP can be understood as a randomized function $\hat{k}$ which is applied to document collections or query results prior to their public release~\cite{dwork2008differential}.
Therefore, for all subsets $S$ in the range of $\hat{k}$, and document collections $D$ and $D'$ differing on at most one element, $\hat{k}$ provides $\epsilon$-differential privacy if:
\begin{equation}\label{eq-1}
    Pr[\hat{k}(D) \in S] \leq exp(\epsilon) Pr[\hat{k}(D') \in S].
\end{equation}
Approaches for DP exploit mathematical formalism to neutralize de-anonymization attacks and keep a lookout for membership inference attacks that may disclose the original documents in the collection. The function $\hat{k}$ adds random noise to any input query and, consequently, yields the response \cite{dwork2008differential}.
Every mechanism that satisfies $\epsilon$-DP will mitigate risks of leakages of private information from any individual element since its inclusion or removal form the collection would not turn the output significantly more or less likely~\cite{dwork2008differential}.
DP provides privacy guarantees usually at the cost of performance and computational overhead.
However, one of the main advantages of this method regards managing the tradeoff between privacy and utility, finding the ideal value of $\epsilon$ which preserves data privacy and affects the model results at a controlled extent.

\subsubsection{ML-based approaches}

\paragraph{Federated learning.}
FL is a methodology for training ML models in distributed computation scenarios proposed by Google~\cite{mcmahan2017communication}.
This methodology prevents data from leaving its owner's device during the computations and addresses privacy threats related to training over private data~\cite{chen2019federated}.
The federated training consists on, firstly, distributing copies of a global model with pre-defined parameters, computing local updates on the distributed devices, sending these updates to the server for aggregation, updating the global model parameters, and sending the updated parameters of the global model to each distributed device~\cite{konevcny2016federated}.
In formal terms, FL assumes a model $\hat{m}$ with parameters $\Theta_{\hat{m}}$ which are stored in a matrix $M$.
The model $\hat{m}$ is thus shared with a subset $T$ of $\eta$ clients, which will update $\hat{m}$ with their locally stored data at each training step $t\geq 0$~\cite{mcmahan2017communication,konevcny2016federated}.
Every client $i$ will send its update $H^{i}_{t} := M^{i}_{t} - M_t$ to the central server, which is responsible for aggregating client-side updates for updating global model~\cite{konevcny2016federated}.
FL has advantages for enabling distributed computations over devices with restrained memory, bandwidth, and computation power.

\section{Discussion}\label{sec:discussion}

A large number of IR systems are sensitive to privacy threats in scenarios which include personal information, search history, personal preferences, private documents, to name a few.
Additionally, regulations as the EU's GDPR establish the guidelines for protecting user generated data from breaches and non-consented usages.
As a consequence, many algorithmic methods for protecting data privacy have been proposed and integrated into IR systems.
However, protecting privacy using such methods can mean coping with utility performance decays and computational overheads.
Therefore, the choice for a convenient privacy-preserving method for an IR scenario has to take into account the target of privacy protection and the computational resources available.

\begin{table}[ht!]
	\caption{Applications of IR alongside suitable privacy-preserving methods.}
	\label{tab:summary-aplications}
	\centering
	\begin{tabular}{l|l}
		\hline
		\textbf{IR application} & \textbf{PP methods}\\
		\hline
		Ad-hoc search & SE, MPC   \\
		Query expansion & HE, SE \\
		Feature extraction for ranking  & DP \\
		Online learning for ranking & DP, HE \\
		Query composition  & SE \\
		Healthcare data tasks & DP, HE, SE \\
		Opinion mining & DP \\
		Advisor for hashtag sharing & SE \\
		Social media profile linking & HE, SE \\
		Recommender systems & FE, MPC \\
		\bottomrule
	\end{tabular}
	
\end{table}

In Table~\ref{tab:summary-aplications}, we bring a summary of IR applications for the privacy-preserving methods surveyed in the section \textit{Solutions}.
Cryptographic approaches, such as HE, SE, and MPC, are suitable to scenarios in which the original content of documents or datasets should not be revealed to unauthorized parties or some of these parties are not trusted, e.g., malicious servers.
DP provides formal privacy guarantees which can be employed on a myriad of applications, such as data anonymization and protecting results of queries and ML models against inversion or reverse engineering.
Finally, FL can be successfully implemented for scenarios with distributed devices and limited resources for data sharing or prohibitions of data exchange.

\section{Conclusion and Future Works}\label{sec:conclusion}

Privacy is a critical point for the development of IR systems which deal with personal, user generated, or sensitive data. 
In this work we overview recent developments in the IR literature, pointing out privacy issues and suggesting suitable privacy-preserving methods.
Data types, IR tasks, and privacy-preserving method drawbacks are taken into account to provide the reader with essential understating of this research field.
As future works, we aim to address the aforementioned challenges for Open Search use cases, as well as studying and discussing compliance with legal requirements, such as those of the EU's GDPR.

\section{ACKNOWLEDGEMENTS}
This work was supported by the EU’s Horizon 2020 project TRUSTS under grant agreement No. 871481.
The Know‐Center is funded within the Austrian COMET Program – Competence Centers for Excellent Technologies – under the auspices of the Austrian Federal Ministry of Transport, Innovation and Technology, the Austrian Federal Ministry of Economy, Family and Youth and by the State of Styria. COMET is managed by the Austrian Research Promotion Agency (FFG).

%
%
\bibliographystyle{IEEEtran}

\begin{thebibliography}{10}
\providecommand{\url}[1]{#1}
\csname url@samestyle\endcsname
\providecommand{\newblock}{\relax}
\providecommand{\bibinfo}[2]{#2}
\providecommand{\BIBentrySTDinterwordspacing}{\spaceskip=0pt\relax}
\providecommand{\BIBentryALTinterwordstretchfactor}{4}
\providecommand{\BIBentryALTinterwordspacing}{\spaceskip=\fontdimen2\font plus
\BIBentryALTinterwordstretchfactor\fontdimen3\font minus
  \fontdimen4\font\relax}
\providecommand{\BIBforeignlanguage}[2]{{%
\expandafter\ifx\csname l@#1\endcsname\relax
\typeout{** WARNING: IEEEtran.bst: No hyphenation pattern has been}%
\typeout{** loaded for the language `#1'. Using the pattern for}%
\typeout{** the default language instead.}%
\else
\language=\csname l@#1\endcsname
\fi
#2}}
\providecommand{\BIBdecl}{\relax}
\BIBdecl

\bibitem{brickley2019google}
D.~Brickley, M.~Burgess, and N.~Noy, ``Google dataset search: Building a search
  engine for datasets in an open web ecosystem,'' in \emph{The World Wide Web
  Conference}, 2019, pp. 1365--1375.

\bibitem{magazine2017landscape}
D.-L. Magazine, ``The landscape of research data repositories in 2015: A
  re3data analysis,'' \emph{D-Lib Magazine}, vol.~23, no. 3/4, 2017.

\bibitem{westin1968privacy}
A.~F. Westin, ``Privacy and freedom,'' \emph{Washington and Lee Law Review},
  vol.~25, no.~1, p. 166, 1968.

\bibitem{EUdataregulations2018}
E.~Commission, ``2018 reform of eu data protection rules,''
  \url{https://ec.europa.eu/commission/sites/beta-political/files/data-protection-factsheet-changes_en.pdf},
  2018, date: 2018-05-25, URL Date: 2019-06-17.

\bibitem{zhang2018tagvisor}
Y.~Zhang, M.~Humbert, T.~Rahman, C.-T. Li, J.~Pang, and M.~Backes, ``Tagvisor:
  A privacy advisor for sharing hashtags,'' in \emph{Proceedings of the 2018
  World Wide Web Conference}, 2018, pp. 287--296.

\bibitem{zhu2017differentially}
T.~Zhu, G.~Li, W.~Zhou, and S.~Y. Philip, ``Differentially private data
  publishing and analysis: A survey,'' \emph{IEEE Transactions on Knowledge and
  Data Engineering}, vol.~29, no.~8, pp. 1619--1638, 2017.

\bibitem{weng2014privacy}
L.~Weng, L.~Amsaleg, A.~Morton, and S.~Marchand-Maillet, ``A privacy-preserving
  framework for large-scale content-based information retrieval,'' \emph{IEEE
  Transactions on Information Forensics and Security}, vol.~10, no.~1, pp.
  152--167, 2014.

\bibitem{tamine2018evaluation}
L.~Tamine and M.~Daoud, ``Evaluation in contextual information retrieval:
  foundations and recent advances within the challenges of context dynamicity
  and data privacy,'' \emph{ACM Computing Surveys (CSUR)}, vol.~51, no.~4, pp.
  1--36, 2018.

\bibitem{gentry2009fully}
C.~Gentry, ``Fully homomorphic encryption using ideal lattices,'' in
  \emph{Proceedings of the forty-first annual ACM symposium on Theory of
  computing}, 2009, pp. 169--178.

\bibitem{cash2015leakage}
D.~Cash, P.~Grubbs, J.~Perry, and T.~Ristenpart, ``Leakage-abuse attacks
  against searchable encryption,'' in \emph{Proceedings of the 22nd ACM SIGSAC
  conference on computer and communications security}, 2015, pp. 668--679.

\bibitem{dwork2008differential}
C.~Dwork, ``Differential privacy: A survey of results,'' in \emph{International
  conference on theory and applications of models of computation}.\hskip 1em
  plus 0.5em minus 0.4em\relax Springer, 2008, pp. 1--19.

\bibitem{feng2020securenlp}
Q.~Feng, D.~He, Z.~Liu, H.~Wang, and K.-K.~R. Choo, ``Securenlp: A system for
  multi-party privacy-preserving natural language processing,'' \emph{IEEE
  Transactions on Information Forensics and Security}, 2020.

\bibitem{konevcny2016federated}
J.~Kone{\v{c}}n{\`y}, H.~B. McMahan, F.~X. Yu, P.~Richt{\'a}rik, A.~T. Suresh,
  and D.~Bacon, ``Federated learning: Strategies for improving communication
  efficiency,'' in \emph{NIPS Work- shop on Private Multi-Party Machine
  Learning}, 2016.

\bibitem{chen2019federated}
M.~Chen, A.~T. Suresh, R.~Mathews, A.~Wong, C.~Allauzen, F.~Beaufays, and
  M.~Riley, ``Federated learning of n-gram language models,'' pp. 121--130,
  2019.

\bibitem{orso2017overlaying}
V.~Orso, T.~Ruotsalo, J.~Leino, L.~Gamberini, and G.~Jacucci, ``Overlaying
  social information: The effects on users’ search and information-selection
  behavior,'' \emph{Information Processing \& Management}, vol.~53, no.~6, pp.
  1269--1286, 2017.

\bibitem{riazi2017prisearch}
M.~S. Riazi, E.~M. Songhori, and F.~Koushanfar, ``Prisearch: Efficient search
  on private data,'' in \emph{2017 54th ACM/EDAC/IEEE Design Automation
  Conference (DAC)}.\hskip 1em plus 0.5em minus 0.4em\relax IEEE, 2017, pp.
  1--6.

\bibitem{mivule2017data}
K.~Mivule, ``Data swapping for private information sharing of web search
  logs,'' \emph{Procedia computer science}, vol. 114, pp. 149--158, 2017.

\bibitem{el2021papir}
A.~El-Ansari, A.~Beni-Hssane, M.~Saadi, and M.~El~Fissaoui, ``Papir:
  privacy-aware personalized information retrieval,'' \emph{Journal of Ambient
  Intelligence and Humanized Computing}, pp. 1--17, 2021.

\bibitem{yang2017differential}
G.~H. Yang and S.~Zhang, ``Differential privacy for information retrieval,'' in
  \emph{Proceedings of the ACM SIGIR International Conference on Theory of
  Information Retrieval}, 2017, pp. 325--326.

\bibitem{dai2020context}
Z.~Dai and J.~Callan, ``Context-aware document term weighting for ad-hoc
  search,'' in \emph{Proceedings of The Web Conference 2020}, 2020, pp.
  1897--1907.

\bibitem{devlin2019bert}
J.~Devlin, M.-W. Chang, K.~Lee, and K.~Toutanova, ``Bert: Pre-training of deep
  bidirectional transformers for language understanding,'' in \emph{Proceedings
  of the 2019 Conference of the North American Chapter of the Association for
  Computational Linguistics: Human Language Technologies, Volume 1 (Long and
  Short Papers)}, 2019, pp. 4171--4186.

\bibitem{esposito2020hybrid}
M.~Esposito, E.~Damiano, A.~Minutolo, G.~De~Pietro, and H.~Fujita, ``Hybrid
  query expansion using lexical resources and word embeddings for sentence
  retrieval in question answering,'' \emph{Information Sciences}, vol. 514, pp.
  88--105, 2020.

\bibitem{carpineto2012survey}
C.~Carpineto and G.~Romano, ``A survey of automatic query expansion in
  information retrieval,'' \emph{Acm Computing Surveys (CSUR)}, vol.~44, no.~1,
  pp. 1--50, 2012.

\bibitem{nogueira2020document}
R.~Nogueira, Z.~Jiang, R.~Pradeep, and J.~Lin, ``Document ranking with a
  pretrained sequence-to-sequence model,'' in \emph{Proceedings of the 2020
  Conference on Empirical Methods in Natural Language Processing: Findings},
  2020, pp. 708--718.

\bibitem{pandey2018linear}
G.~Pandey, Z.~Ren, S.~Wang, J.~Veijalainen, and M.~de~Rijke, ``Linear feature
  extraction for ranking,'' \emph{Information Retrieval Journal}, vol.~21,
  no.~6, pp. 481--506, 2018.

\bibitem{zhuang2020counterfactual}
S.~Zhuang and G.~Zuccon, ``Counterfactual online learning to rank,'' in
  \emph{European Conference on Information Retrieval}.\hskip 1em plus 0.5em
  minus 0.4em\relax Springer, 2020, pp. 415--430.

\bibitem{chuklin2015click}
A.~Chuklin, I.~Markov, and M.~d. Rijke, ``Click models for web search,''
  \emph{Synthesis lectures on information concepts, retrieval, and services},
  vol.~7, no.~3, pp. 1--115, 2015.

\bibitem{zolaktaf2020facilitating}
Z.~Zolaktaf, M.~Milani, and R.~Pottinger, ``Facilitating sql query composition
  and analysis,'' in \emph{Proceedings of the 2020 ACM SIGMOD International
  Conference on Management of Data}, 2020, pp. 209--224.

\bibitem{sun2018data}
W.~Sun, Z.~Cai, Y.~Li, F.~Liu, S.~Fang, and G.~Wang, ``Data processing and text
  mining technologies on electronic medical records: a review,'' \emph{Journal
  of healthcare engineering}, vol. 2018, 2018.

\bibitem{nguyen2018multilingual}
H.~T. Nguyen and M.~Le~Nguyen, ``Multilingual opinion mining on youtube--a
  convolutional n-gram bilstm word embedding,'' \emph{Information Processing \&
  Management}, vol.~54, no.~3, pp. 451--462, 2018.

\bibitem{hadgu2020learn2link}
A.~T. Hadgu and J.~K.~R. Gundam, ``Learn2link: Linking the social and academic
  profiles of researchers,'' in \emph{Proceedings of the International AAAI
  Conference on Web and Social Media}, vol.~14, 2020, pp. 240--249.

\bibitem{wang2015collaborative}
H.~Wang, N.~Wang, and D.-Y. Yeung, ``Collaborative deep learning for
  recommender systems,'' in \emph{Proceedings of the 21th ACM SIGKDD
  international conference on knowledge discovery and data mining}, 2015, pp.
  1235--1244.

\bibitem{qi-etal-2020-privacy}
\BIBentryALTinterwordspacing
T.~Qi, F.~Wu, C.~Wu, Y.~Huang, and X.~Xie, ``Privacy-preserving news
  recommendation model learning,'' in \emph{Findings of the Association for
  Computational Linguistics: EMNLP 2020}.\hskip 1em plus 0.5em minus
  0.4em\relax Online: Association for Computational Linguistics, Nov. 2020, pp.
  1423--1432. [Online]. Available:
  \url{https://www.aclweb.org/anthology/2020.findings-emnlp.128}
\BIBentrySTDinterwordspacing

\bibitem{liu2020mitigating}
H.~Liu and B.~Wang, ``Mitigating file-injection attacks with natural language
  processing,'' in \emph{Proceedings of the Sixth International Workshop on
  Security and Privacy Analytics}, 2020, pp. 3--13.

\bibitem{dai2019efficient}
X.~Dai, H.~Dai, G.~Yang, X.~Yi, and H.~Huang, ``An efficient and dynamic
  semantic-aware multikeyword ranked search scheme over encrypted cloud data,''
  \emph{IEEE Access}, vol.~7, pp. 142\,855--142\,865, 2019.

\bibitem{cramer_damgard_nielsen_2015}
R.~Cramer, I.~B. Damgård, and J.~B. Nielsen, \emph{Secure Multiparty
  Computation and Secret Sharing}.\hskip 1em plus 0.5em minus 0.4em\relax
  Cambridge University Press, 2015.

\bibitem{mcmahan2017communication}
B.~McMahan, E.~Moore, D.~Ramage, S.~Hampson, and B.~A. y~Arcas,
  ``Communication-efficient learning of deep networks from decentralized
  data,'' in \emph{Artificial Intelligence and Statistics}.\hskip 1em plus
  0.5em minus 0.4em\relax PMLR, 2017, pp. 1273--1282.

\end{thebibliography}

%
%


\end{document}